\documentclass[a4paper]{article}

\usepackage{amsmath}
\usepackage{algorithm}
\usepackage[noend]{algpseudocode}
\makeatletter
\def\BState{\State\hskip-\ALG@thistlm}
\makeatother

\usepackage{booktabs,caption}
\usepackage[flushleft]{threeparttable}

\usepackage{hhline}

\usepackage{makecell}

\usepackage{enumitem}

\usepackage{INTERSPEECH2018}

\usepackage{xcolor}
\definecolor{Hc4}{cmyk}{0, 1, 0.63,  0.49}

\newcommand{\Pg}[1]{\noindent\paragraph*{{\color{Hc4}#1}}}

\renewcommand{\Pg}[1]{}

\newcommand{\ie}{\textit{i.e.,\ }}
\newcommand{\etal}{\textit{et al.\@}}

\title{Multimodal Speaker Segmentation and Diarization using Lexical and Acoustic Cues via Sequence to Sequence Neural Networks}
\name{Tae Jin Park, Panayiotis Georgiou}
\address{
  University of Southern California, Los Angeles, CA, USA }
\email{taejinpa@usc.edu, georgiou@sipi.usc.edu}

\begin{document}

\maketitle
\begin{abstract}

  While there has been substantial amount of work in speaker diarization recently, there are few efforts in jointly employing lexical and acoustic information for speaker segmentation.
  Towards that, we investigate a speaker diarization system using a sequence-to-sequence neural network trained on both lexical and acoustic features.
  We also propose a loss function that allows for selecting not only the speaker change points but also the best speaker at any time by allowing for different speaker groupings.
  We incorporate Mel Frequency Cepstral Coefficients (MFCC) as an acoustic feature alongside lexical information that are obtained from conversations from the Fisher dataset. 
Thus, we show that acoustics provide complementary information to the lexical modality. 
The experimental results show that sequence-to-sequence system trained on both word sequences and MFCC can improve on speaker diarization result compared to the system that only relies on lexical modality or the baseline MFCC-based system. 
In addition, we test the performance of our proposed method with Automatic Speech Recognition (ASR) transcripts. 
While the performance on ASR transcripts drops, the Diarization Error Rate (DER) of our proposed method still outperforms the traditional method based on Bayesian Information Criterion (BIC). 

\end{abstract}



\section{Introduction}

Speaker Diarization is an important pre-processing step towards a complete Automatic Speech Recognition (ASR) system that includes multiple speakers.
Further, speaker diarization information plays crucial a role in speech analytics such as turn-taking characteristics and is critical in many behavioral analytics applications \cite{georgiou2011_behavioral-sign, narayanan2013_behavioral-sign}.
Poor performance of speaker diarization is bound to deteriorate the performance of subsequent models such as ASR, emotion recognition, behavioral informatics, and topic analysis systems.
Speaker segmentation is a critical component of this process and heavily affects the performance of speaker diarization and hence all subsequent modules.

In general, a speaker diarization system consists of two main parts: segmentation and clustering. Segmentation aims to detect all speaker change points.
The most widely used method is the Bayesian Information Criterion (BIC) based segmentation \cite{tritschler1999improved, chen1998speaker}.
More recently, methods based on Recursive Neural Networks (RNN) have shown improved performance on speaker segmentation \cite{yin2017speaker, wang2017speaker}.
In addition, Joint Factor Analysis (JFA) \cite{desplanques2015factor} has also shown promising results.
Further, there are significant efforts in speaker segmentation and diarization with pre-trained Deep Neural Networks (DNN) both through supervised-training \cite{garcia2017speaker} and through unsupervised-training \cite{jati2017speaker2vec,jati2018-neural-predicti}.

Despite the very active field, there has been very little effort in exploiting lexical information towards this task.
Most of the research that involves lexical information or transcript is relating to speaker identity \cite{canseco2004speaker, esteve2007extracting} or speaker role \cite{xiao2015_rate-my-therapi,xiao2016_a-technology-pr}.
India \etal\ employed character level information via an LSTM network with a character level Convolutional Neural Network (CNN) and i-vector training on transcript \cite{india2017lstm}.

One likely reason that transcripts from ASR have not been used for diarization is that we often are hesitant to run ASR before diarization since that will be more noisy that employing these two components in reverse order.
However that is not a constraint (except in computation resources) as the ASR can be re-run after diarization a second time.
Further, along recent efforts of research including in our group, of joint training, future implementations can jointly optimize for diarization and ASR.

In our work we \emph{propose} a system that incorporates both lexical cues and acoustic cues to build a system closer to how humans employ information. We investigate a sequence-to-sequence model (seq2seq) that  integrates both lexical  and acoustic cues to perform speaker segmentation and speaker diarization.
Sequence-to-sequence models have been widely used for language translation \cite{sutskever2014sequence}, end to end ASR systems \cite{chan2016listen} and text summarization\cite{nallapati2016abstractive}.
The advantage of seq2seq over Recurrent Neural Network (RNN) based models (LSTM \cite{hochreiter1997long}, GRU \cite{cho2014learning}) is that it  can summarize the whole sequence into an embedding and then pass it to the decoder.
Moreover, it can integrate information and process variable length sequences.
In doing so, such a  model can capture temporally encoded information from both before and after the speaker change points.
In addition, the attention mechanism of this model helps in capturing the important parts of characterizing the speaker(s).

In our work we employ dyadic-interaction data to train and test the proposed system. 
Our proposed model operates on both reference transcript data and, critically for realistic deployment,  on ASR hypotheses.

\section{Proposed Speaker Diarization System}
\subsection{Network Architecture}
Our proposed sequence to sequence model consists of encoder, decoder and attention model that connects encoder and decoder.
The encoder consumes a sequence of word representations, along with acoustic features (MFCC) described in sec.~\ref{sec:features}, as shown in Fig.~\ref{fig:encoder_fig}.
The decoder produces a sequence of words along with speaker IDs during the speaker change points, as shown in Fig.~\ref{fig:decoder_fig}.
We used GRU with a 256-dimensional hidden layer and an attention model that has been applied to many state-of-the-art machine translation systems \cite{bahdanau2014neural}. 

\subsection{Feature processing}\label{sec:features}

In our proposed method the features are time-synchronous. All the features align with the word boundaries as follows:
\begin{description}
\item[WORD:] The word sequences we use are obtained either from the reference transcripts or from an ASR output. We use a linear layer to convert one hot word vector into word embedding as described in Fig.~\ref{fig:encoder_fig}. The source sequence is 32 words in the reference transcript or ASR output. The target sequence for training is 32 words and added speaker turn tokens as in the example sentence in table\ref{example_data}. 
\item[MFCC:] We used 13-dimensional MFCCs extracted with a 25ms window and 10ms shift. 
 Detailed specifications follow the default settings in \cite{lyons2017python}. We then average the MFCC features for the word-segment and thus derive  a 13 $\times$ 1 vector for each word.
\end{description}

\subsection{Encoder and input features}
\begin{figure}[t]
  \centering
  \centerline{\includegraphics[width=0.8\columnwidth]{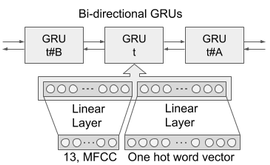}}
  \caption{The encoder side of the proposed network.}
  \label{fig:encoder_fig}
\end{figure}

In our proposed system, the encoder integrates MFCC feature vector and word embedding. Fig.\ref{fig:encoder_fig} shows how the proposed encoder is structured. Word embeddings, MFCC and pitch features are connected through linear layers. After the fully-connected layers, the embeddings are concatenated. The concatenated vector is then fed to the GRU that is the  encoder of the seq2seq system. We use 256 hidden unit size, word embedding size and output layer of linear layer for MFCC vector.
The number of hidden layers were chosen to be equal for both MFCC and word embedding because there is a performance degradation when these embedding size are different. However, more optimization needs to take place for the most optimal system. 
\subsection{Decoder and loss function}
\begin{figure}[t]
  \centering
  \includegraphics[width=\linewidth]{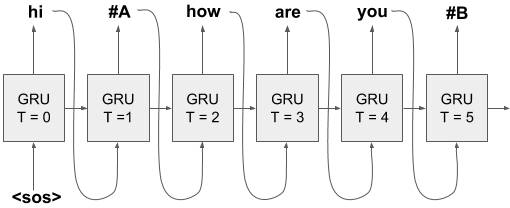}
  \caption{The decoder side of the proposed sequence to sequence model.}
  \label{fig:decoder_fig}
  \vspace{-1.0ex}
\end{figure}

In our proposed system, the decoder outputs a word sequence and the speaker turn token ``$\sharp$A'' and ``$\sharp$B''. Fig.~\ref{fig:decoder_fig} describes the decoder side in our proposed system. Unlike word tokens, the loss of the speaker turn tokens are calculated in a different way that ignores the speaker IDs and only focuses on speaker groupings. For example, the speaker turn sequence of ``$\sharp$A $\sharp$B $\sharp$A'' is considered equal to ``$\sharp$B $\sharp$A $\sharp$B''. That is, the loss function in our proposed system calculates two versions of losses: original and flipped version of speaker turn tokens. Between these two losses, our loss function selects the smaller loss. 
This loss function also  avoids learning the probability between speaker turn tokens and words in the target sequences in the training set.

\begin{table}[t]
\centering
\caption{An example of source sentence and target sentence in training data.}
\label{example_data}
\begin{tabular}{l|l}
 \hline
\textbf{Source} & hello hi my name is James hi James  \\ \hline
\textbf{Target} & hello $\sharp$A hi $\sharp$B my name is James $\sharp$A hi James \\  \hline   
\end{tabular} 
\vspace{-1.0ex}
\end{table}

\subsection{Speaker Turn Estimation}
\label{sec:spkt}
\begin{figure}[t]
  \centering
  \centerline{\includegraphics[width=0.95\columnwidth]{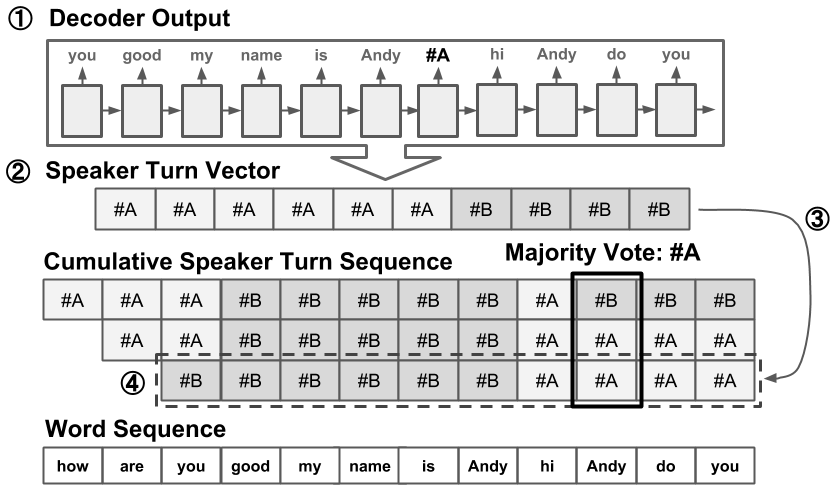}}
  \caption{Decoder output and overlapping speaker turn vectors.}
  \label{fig:spkt_vector}
  \vspace{-3.0ex}
\end{figure}

To maximize the accuracy of speaker turn detection, we employ shift and overlap scheme to predict the speaker turn. Fig.~\ref{fig:spkt_vector} explains how speaker turn prediction is done. A target window that has 32 word length sweeps the whole session from the beginning to the end. For each target window, we predict speaker turn tokens with our trained sequence to sequence model. At each prediction, we extract 32 words and 32 MFCC vectors from transcript and audio stream, respectively. A set of speaker turns for a session is estimated through the following process in accordance with the indices in Fig,~\ref{fig:spkt_vector}.
\begin{description}[labelindent=0cm, leftmargin=0cm]
\item[1.] Obtain a new word sequence and estimated speaker turn tokens from decoder outputs.
\item[2.] Form a speaker turn vector by assigning each word the nearest speaker turn token.
\item[3.] Store the speaker turn vector that is obtained from step 2 in a cumulative speaker turn sequence which is the  matrix that sequentially stores all the speaker turn vectors obtained so far. Flip the speaker turn vector if flipping the speaker turn vector gives less hamming distance with all the other speaker turn tokens in cumulative speaker turn sequence.
\item[4.] Store the speaker turn vector from step 3 into the cumulative speaker turn sequence. Shift one word to the right and feed next 32 words and 32 MFCC vectors to the encoder of the proposed system. 
\end{description}
After finishing the above process by shifting 32 word window to the end of the session, we determine the final speaker turn decision by taking a majority vote. In this way, a word in a session incorporates 32 different predictions to determine the speaker turn.

\subsection{Clustering}
We will evaluate on diarization accuracy we therefore employ our SCUBA, BIC based agglomerative clustering algorithm based on \cite{chen1998speaker} to perform the clustering step. For the agglomerative clustering we employ the raw frame-level MFCC  as features. We obtain the segmented MFCC streams using speaker turn information that is produced from the process described in~\ref{sec:spkt}. This clustering algorithm is applied to all of the models in this paper, including the LIUM baseline. For the baseline systems, the process mentioned in \ref{sec:spkt} is replaced with other methods while same agglomerative clustering algorithm is applied. 

\section{Experimental Results}

Our proposed system is tested with two different datasets: those stemming from reference transcription and those from automatically derived ASR hypotheses.

\Pg{Training data:}
To train our proposed system with dialogue, we train our proposed system on Fisher English Training Speech Part 1 and Part 2 \cite{cieri2004fisher} for both lexical cues and acoustic cues. This results in 11,112 training dialogs comprised of  approximately 19 million words.

\Pg{Evaluation data:} Before training the proposed system, we randomly chose and separated 20 sessions as a test set and 567 sessions as a dev-set from the original Fisher dataset. These are used as evaluation in the case we employ clean transcripts.
For evaluation using ASR output, we also use Switchboard-1 Telephone Speech Corpus \cite{godfrey1997switchboard} to ensure complete train-test separation and domain generalization.
\Pg{Oracle error:}
Although the original recordings were 2-channel telephony (1 per speaker) we generate single channel signals by mixing down to mono. For the word alignment information, forced-alignment was used to obtain the word alignment information for Fisher dataset since word-level alignment information is not provided in Fisher dataset while speaker turn level alignment is provided. For Switchboard-1 dataset, we use provided word alignment information and speaker turn level alignment information. With this alignment information, we create the ground truth diarization labels for subsequent evaluation. Due to the overlaps in the data the lower-bound diarization error is not zero, and we will thus also denote that in the tables below.
\Pg{Baseline:}

As a benchmark of our proposed method we employ LIUM Speaker Diarization Tools \cite{rouvier2013open} which contains a Speaker Activity Detection (SAD) system and a speaker segmentation system. The LIUM script that we use performs MFCC feature extraction, SAD and speaker segmentation sequentially. We used default settings for all the parameters. The clustering step is employing the same algorithm as all other methods in this paper (\ie LIUM segmentation and SCUBA clustering)

A second baseline is to employ agglomerative clustering for diarization but by employing the word boundaries as segmentation. For convenience, we refer to this model as WS. WS baseline can verify the merit of our proposed model since we can compare whether the performance is stemming from word alignment or speaker turn probability when we estimate with our proposed system. For reference transcript based test, WS is obtained from word alignment data in the transcript and for ASR transcript based test, WS is obtained from word alignment data from ASR transcript.
\Pg{Metrics:}
 We are using Diarization Error Rate (DER) as a performance metric for all experiments. To measure the DER metric, we employ the \textit{md-eval} software in RT06S dataset \cite{fiscus2006rich} with the forgiveness collar of 0.25 seconds.

\begin{figure}[t]
  \centering
  \centerline{\includegraphics[width=0.95\columnwidth]{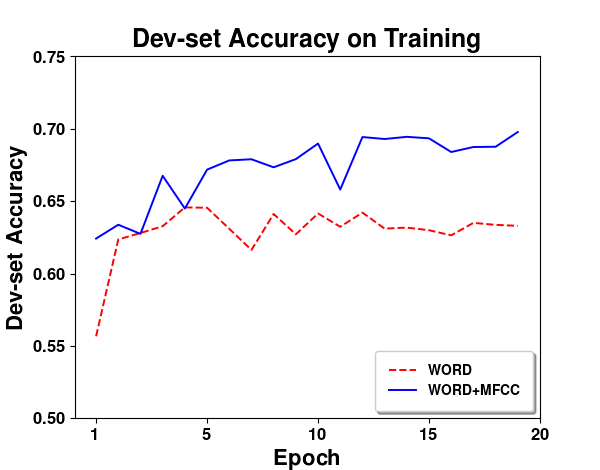}}
  \caption{Dev-set accuracy on training. }
  \label{fig:train_acc}
  \vspace{-3.0ex}
\end{figure}
\subsection{Training of sequence to sequence model}
We train and test three different models separately. Each model employs the same architecture and the same training conditions except the feature.
The first model is trained only on word embeddings, the second one is trained on both word embeddings and MFCC. For convenience, we will refer to these as W model and WM model respectively. We train each model until convergence (20 epochs). We use teacher forcing \cite{williams1989learning} ratio of 0.5 to speedup training. Fig.~\ref{fig:train_acc} shows the dev-set accuracy while training. The WM model clearly shows improved performance over W model. Note that accuracy in Fig.~\ref{fig:train_acc} is accuracy measured with word sequence that contains speaker turn tokens and word tokens. Thus, this accuracy does not always mean better segmentation or diarization accuracy.  

\subsection{Experiment on Reference Transcripts}

First we experiment using reference transcripts. In this case MFCC features are obtained using the oracle word alignments. Thus, we use accurate word embedding and temporal information of each word. Table \ref{der_trans} shows the result we obtained from transcript data.

The result clearly shows that incorporating  MFCC features helps the performance of diarization when the word embedding and temporal information is accurate. In addition, W model and WM model also outperformed word-level segmentation (WS) based result. This suggests that applying our proposed model gives a merit over simply using word-alignment information as segmentation result. 
We also tested the diarization system with ground truth speaker label per word and it showed 16.22\% and 18.06\% for Fisher and Switchboard data respectively. This is due to the frequent overlaps in dialogues and inaccurate labeling of speaker turn level transcript data. Therefore, ``Oracle'' DER  in table \ref{der_trans} is the best  performance we can achieve with any algorithm. To check the performance of the proposed system in different way, we also measured Word-level Diarization Error Rate (WDER) which means ``who says this word''. Table \ref{wder_trans} shows WDER result for transcript based experiment. Since there are two speakers in this experiment, the WDER also shows similar result to DER result where WM model shows nearly 4\% improvement over W model.

\subsection{Experiment on ASR transcript}
For ASR transcript, we use the Kaldi Speech Recognition Toolkit \cite{povey2011kaldi} and ASR model trained on whole Fisher English Speech data. As a test-set, we chose the 30 audio files with  lowest index ID in each of the 30 folders of the Switchboard-1 dataset for reproducibility of our experiment. Table \ref{der_asr} shows the result from ASR based experiment. Unlike in the case of reference transcripts, in this case WM model did not improve the performance. However, ASR based result is still better than diarization based on segmentation result obtained from LIUM Speaker Diarization Tools. In addition, WS model also performed better than LIUM Tools, which indicates using word-level segmentation from ASR can still perform better than BIC based segmentation system.

For ASR transcript, we use the Kaldi Speech Recognition Toolkit \cite{povey2011kaldi} and ASR model trained on whole Fisher English Speech data. As a test-set, we choose the 30 audio files that have lowest index in each of 30 folders in Switchboard-1 dataset for reproducibility of our experiment. Table \ref{der_asr} shows the result from ASR based experiment. Unlike in the case of reference transcripts, WM model did not improve the performance. However, ASR based result is still better than diarization based on segmentation result obtained from LIUM Speaker Diarization Tools. In addition, WS model also performed better than LIUM Speaker Diarization Tools, which indicates using word-level segmentation from ASR can still perform better than BIC based segmentation system.  

\subsection{WER vs DER}
\begin{figure}[t]
  \centering
  \centerline{\includegraphics[width=0.95\columnwidth]{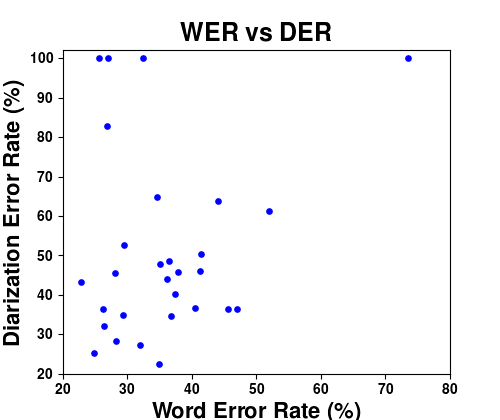}}
  \caption{Scatter plot of WER vs DER}
  \label{fig:wer_der}
\end{figure}

Since we test the improvement by incorporating acoustic cues with transcript data, performance degradation in the experiment with ASR transcript is solely caused by poor ASR Word Error Rate (WER). The average WER for 30 Switchboard  session is 35.15\%. Fig.~ \ref{fig:wer_der} shows the scatter plot between WER vs DER for the experiment with ASR transcript (Table \ref{der_asr}). As we can see in Fig.~\ref{fig:wer_der}, no session shows low DER when WER is high. However, although WER is pretty low, DER can be very high. Based on this outcome, we could conclude that low WER is necessary condition for low DER, not the sufficient condition.

\begin{table}[t]
\centering
\caption{DER on transcription data.}
\label{der_trans}
\begin{tabular}{c|c c c c || c }
 \hline \vspace{0.3ex}
DER(\%) &   \textbf{W}  & \textbf{WM} & \textbf{WS} & Oracle & \textbf{LIUM} \\ \hline
Fisher     & 28.02 & 24.26     & 44.53     & 16.22 & 77.45 \\
Switchboard & 27.89 & 22.44     & 46.4 & 18.06 & 66.57\\ \hline
\end{tabular}

\end{table}

\begin{table}[t]
\centering
\caption{WDER on transcription data.}
\label{wder_trans}
\begin{tabular}{c|cccc}
\hline
WDER(\%)     & \textbf{W}  & \textbf{WM}   \\ \hline
\makecell{Fisher \\ Transcript}     & 16.42 & 12.32      \\ \hline
\makecell{Switchboard \\ Transcript} & 12.4 & 8.56     \\ \hline          
\end{tabular}
\end{table}

\begin{table}[t]
\centering
\caption{DER on ASR transcript and baseline system.}
\label{der_asr}

\begin{tabular}{c|cccc||c}
\hline
DER(\%)     & \textbf{W}  & \textbf{WM}  & WS  & Oracle & \textbf{LIUM}\\ \hline
\makecell{Switchboard \\ASR} & 38.64 & 50.95 & 46.02 & 18.06 & 66.57 \\ \hline
\end{tabular} 

\end{table}

\section{Discussion}
Comparing the two experiments using the reference transcripts and ASR transcripts with our proposed system shows that ASR performance hugely affects the performance of DER. However, the experiment with transcript still shows that acoustic cues can improve the diarization performance. Therefore, we can conclude that acoustic cues can be integrated with lexical cues but the ASR performance is critical.
Further we believe that many of the errors that are made by the ASR in segmentation step may create unrecoverable errors, and hence this points to potential benefits of using lattice information and exploiting the ASR uncertainty.


\section{Conclusions}
In this paper, we investigated the way to integrate lexical cues and acoustic cues with sequence to sequence model to improve speaker diarization performance. The  results show very strong support that lexical information can improve the speaker diarization system. We also see that ASR performance plays a crucial role to the performance of our proposed system and poor WER degrades the proposed system trained on both acoustic features and word embeddings. The future work might include improving performance by training data on ASR transcript including multiple-hypotheses to provide alternate word alignment and segmentation points. Further we will investigate use of alternate  acoustic feature representations such as i-vector or embeddings obtained from neural networks\cite{jati2018-neural-predicti,jati2017speaker2vec}. In addition, fusion of frame and word level segmentation will also be considered to increase flexibility on segmentation decisions.

\section{Acknowledgements}
\label{sec:ack}
The U.S. Army Medical Research Acquisition Activity, 820 Chandler Street, Fort Detrick MD 21702-5014 is the awarding and administering acquisition office. This work was supported by the Office of the Assistant Secretary of Defense for Health Affairs through the Psychological Health and Traumatic Brain Injury Research Program under Award No. W81XWH-15-1-0632. Opinions, interpretations, conclusions and recommendations are those of the author and are not necessarily endorsed by the Department of Defense.

\bibliographystyle{IEEEtran}
\bibliography{mybib}{}


\end{document}